\documentclass[aps,12pt,tightenlines,nofootinbib,showpacs,showkeys,eqsecnum]{revtex4}
\usepackage{amsmath,amscd,amssymb}
\usepackage{color,epsfig}

\def\be{\begin{equation}}
\def\ee{\end{equation}}
\def\bea{\begin{eqnarray}}
\def\eea{\end{eqnarray}}

\newcommand{\fract}[2]{\mbox{\small $\frac{#1}{#2}$}}
\newcommand{\half}{\fract{1}{2}}
\newcommand{\zr}[1]{\mbox{\hspace*{#1em}}}
\newcommand{\ID}{\mbox{{\sf 1}\zr{-0.16}\rule{0.04em}{1.55ex}\zr{0.1}}}
\newcommand{\eq}[1]{eq.~(\ref{#1})}

\newcommand{\ie}{{\it i.e.\@}~}
\newcommand{\eg}{{\it e.g.\@}~}
\newcommand{\cf}{{\it cf.\@}~}

\begin{document}

\title{Two--photon contributions to the Rosenbluth cross--section \\
in the Skyrme model}

\author{M.\@ Kuhn and H.\@ Weigel}

\affiliation{Fachbereich Physik, Siegen University,  
D--57068 Siegen, Germany}

\begin{abstract}
We study two--photon contributions to the elastic electron nucleon 
scattering within the Skyrme model. In particular we focus on the
role of the anomaly that enters via the Wess--Zumino term and explain
how this induces an axial current interaction.
\end{abstract}


\maketitle

\section{Introduction}

Since decades the so--called Rosenbluth separation has been utilized to extract 
the electromagnetic form factors of the nucleon from the differential cross 
section for unpolarized electron nucleon scattering. More recently, polarization
measurements have become available that provide additional information on
these form factors. Surprisingly, substantial inconsistencies between these 
two measurements of the same physical quantity seem to emerge. 

To start discussing the problem we introduce the relevant Lorentz invariant 
kinematical variables. First we have 
\be
\tau =\frac{Q^2}{4M^2}=-\frac{q^2}{4M^2}
\label{eq:deftau}
\ee
where $q=k-k^\prime=p^\prime-p$ is the momentum transfer and $M$ the nucleon 
mass. For space--like processes, such as the elastic electron nucleon scattering,
$Q^2=-q^2$ is non--negative. The second Lorentz invariant variable is the 
photon polarization parameter
\be
\epsilon=\frac{\nu^2-M^4\tau\left(1+\tau\right)}
{\nu^2+M^4\tau\left(1+\tau\right)}
\qquad \mbox{where}\qquad
\nu=\frac{1}{4}\left(k+k^\prime\right)\cdot\left(p+p^\prime\right)\,.
\label{eq:defeps}
\ee

\begin{figure}[t]
\centerline{
\epsfig{file=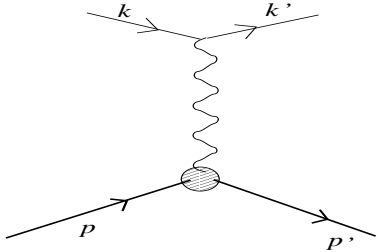,width=5cm,height=3.3cm}}
\caption{\label{fig:leading}
Tree level loop contribution to the electron (momenta $k$ and $k^\prime$)
nucleon momenta $p$ and $p^\prime$) scattering with a local nucleon two--photon 
vertex.}
\end{figure}

The leading (tree level) contribution to elastic electron nucleon scattering
is shown as Feynman diagram in figure~\ref{fig:leading}. In this one--photon 
exchange (or Born) approximation the unpolarized elastic electron
nucleon scattering cross section is
\be
\frac{d\sigma}{d\Omega}=\left(\frac{d\sigma}{d\Omega}\right)_{\rm Mott}
\hspace{-0.5cm}
\frac{\epsilon\, G_E^2(Q^2)+\tau\,G^2_M(Q^2)}{\epsilon(1+\tau)}
\label{eq:Rbluth}
\ee
in the limit of vanishing electron mass, $m_e\to0$, which is well justified
in the considered kinematical regime. In \eq{eq:Rbluth}
$G_{E,M}$ are the electromagnetic form factors of the nucleon
to be further specified later. The Rosenbluth separation to extract these
form factors from data is to display the ratio of the measured cross 
section and the Mott cross section at a given momentum transfer, \ie $\tau$ as 
function of $\epsilon$. After multiplicating this function with
$\epsilon(1+\tau)$, the intercept and slope yield $G_M$ and $G_E$, respectively. 

However, since the relative contribution 
of $G_E$ to the cross section quickly decreases with growing momentum transfer, 
this type of analysis becomes increasingly difficult already at $Q^2\sim 1{\rm GeV}^2$. 
Yet the results of this separation have turned out consistent with the 
assumption that the ratio
\be
R(Q^2)=\frac{\mu_p G_E(Q^2)}{G_M(Q^2)}\,,
\label{eq:ratio}
\ee
where $\mu_p=G_M(0)$ is the proton magnetic moment, approximately equals 
unity~\cite{HWA}. More recently, it has become possible to directly extract
this ratio from polarization observables~\cite{JGP}, thereby avoiding the 
potentially uncertain separation technique. Most surprisingly, these direct
measurements exhibited a linear fall 
\be
R(Q^2)\approx1-0.13\left(Q^2[\rm GeV]^2-0.04\right)
\label{eq:Rpol}
\ee
indicating an eventual
root at about $Q^2\sim10{\rm GeV}^2$. Such a behavior was already suggested
as early as 1973~\cite{Iachello:1972nu} within semi--empirical fits 
to existing data. More recently this structure has been predicted 
within a chiral soliton model study~\cite{Ho}. To resolve this puzzle, the
Rosenbluth separation has been repeated with significantly improved 
precision~\cite{CQ}. For this technique to be operative, it is important that 
the data are indeed consistent with this linear relation~\cite{TGT}. Not only
is this indeed the case but also the previous result $R\approx 1$ is 
reproduced. Hence we face the paradox situation that two distinct methods
to experimentally determine a fundamental nucleon property yield 
inconsistent results~\cite{AG}.

A possible resolution could be that contributions to the cross section
that stem from two--photon exchanges but have been omitted so far, are 
amplified with increasing $Q^2$. In turn, the Rosenbluth separation
yields modified form factors that significantly deviate from the ones
that are defined via a one photon exchange~\cite{Guichon:2003qm}. On the 
other hand, symmetry properties require that the two--photon exchange 
contributions alter the (linear) dependence on the photon polarization
parameter $\epsilon$~\cite{Rekalo:2003xa}. Yet the data are consistent 
with but not restricted to this linear dependence~\cite{Chen:2007ac}.
It is widely believed~\cite{Carlson:2007sp} that the two--photon 
mainly effects the Rosenbluth method but are negligibly small for the 
polarization process. We will therefore focus on the former.
\begin{figure}[t]
\centerline{
\epsfig{file=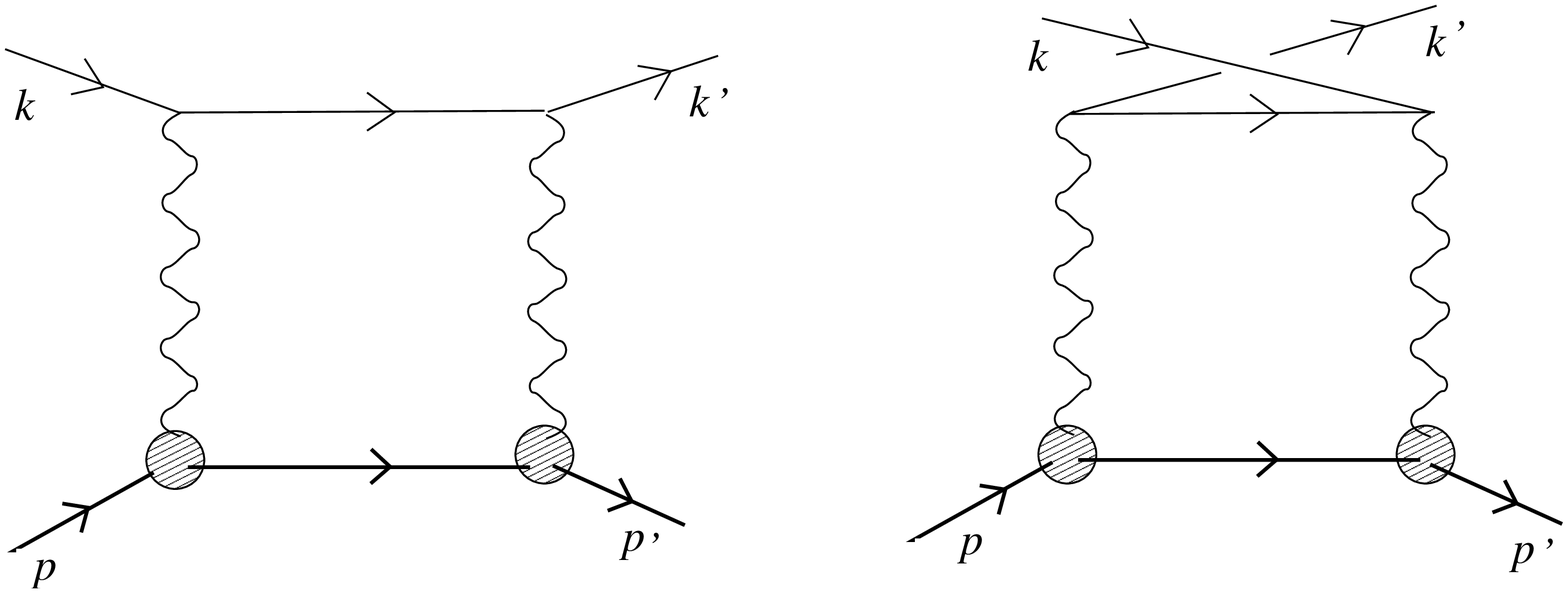,width=10cm,height=3.5cm}}
\caption{\label{fig:box}
One loop contribution to the electron (momenta $k$ and $k^\prime$) nucleon 
(momenta $p$ and $p^\prime$) scattering within a typical box diagram. In the 
hadronic picture, the intermediate baryon can be any nucleon 
resonance~\cite{Kondratyuk:2007hc}, while in the quark picture the photon nucleon 
vertex can, \eg be related to generalized parton distributions~\cite{Carlson:2007sp}.}
\end{figure}

\begin{figure}[t]
\centerline{
\epsfig{file=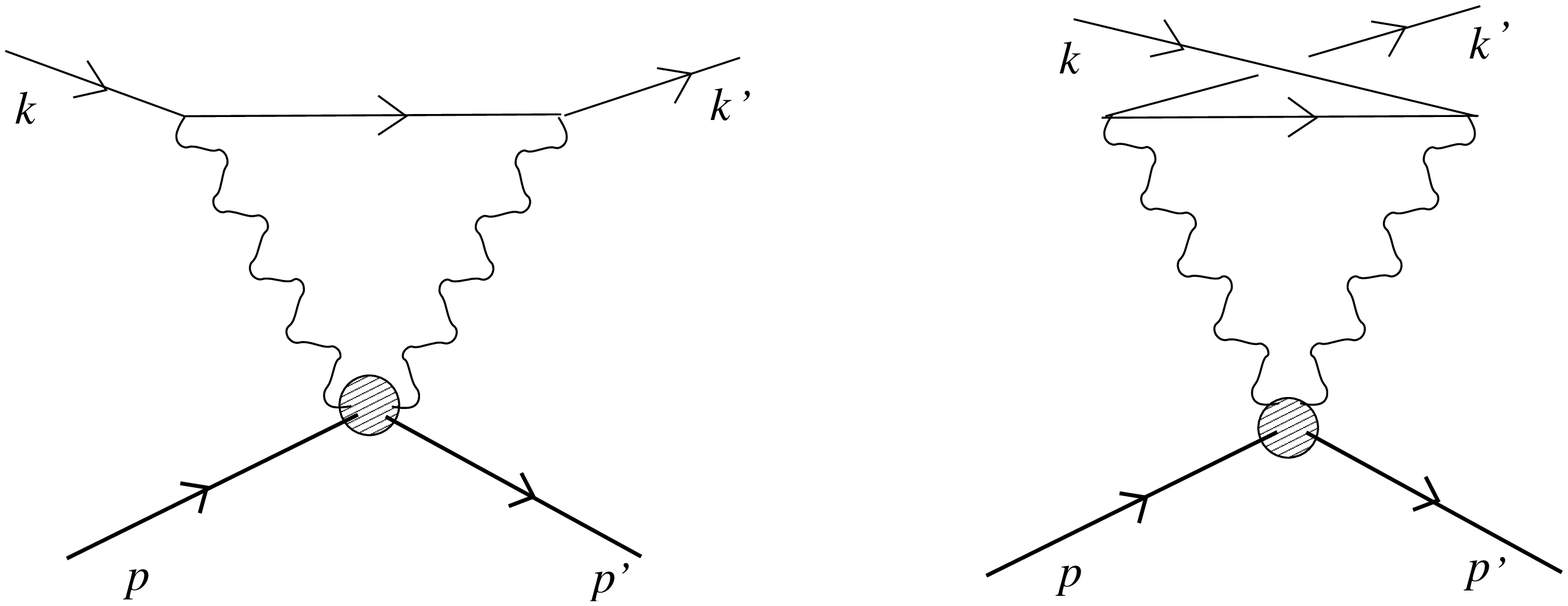,width=10cm,height=3.5cm}}
\caption{\label{fig:2ph1}
One loop contribution to the electron (momenta $k$ and $k^\prime$) nucleon 
(momenta $p$ and $p^\prime$) scattering with a local nucleon two--photon vertex.}
\end{figure}

There are two types of two--photon exchange Feynman diagrams that we 
display in figures~\ref{fig:box} and~\ref{fig:2ph1}. The box diagrams 
in figure~\ref{fig:box} are essentially iterations of the one--photon
exchange shown in figure~\ref{fig:leading}. Any estimate of these box 
diagrams require additional information about the off--shell behavior of 
the photon--nucleon vertex. Furthermore the intermediate baryon is not
restricted to be the nucleon, it could be any nucleon resonance that 
possesses a sizable electro--production potential. Many assumptions and 
modeling of baryon properties enter the computation of these
diagrams~\cite{Guichon:2003qm,Blunden:2005ew,Jain:2006mu,Arrington:2007ux,Kondratyuk:2007hc}. 
For a recent review on both the experimental and the theoretical
situations see ref.~\cite{Perdrisat:2006hj,Carlson:2007sp}.

\begin{figure}[t]
\begin{center}
\epsfig{file=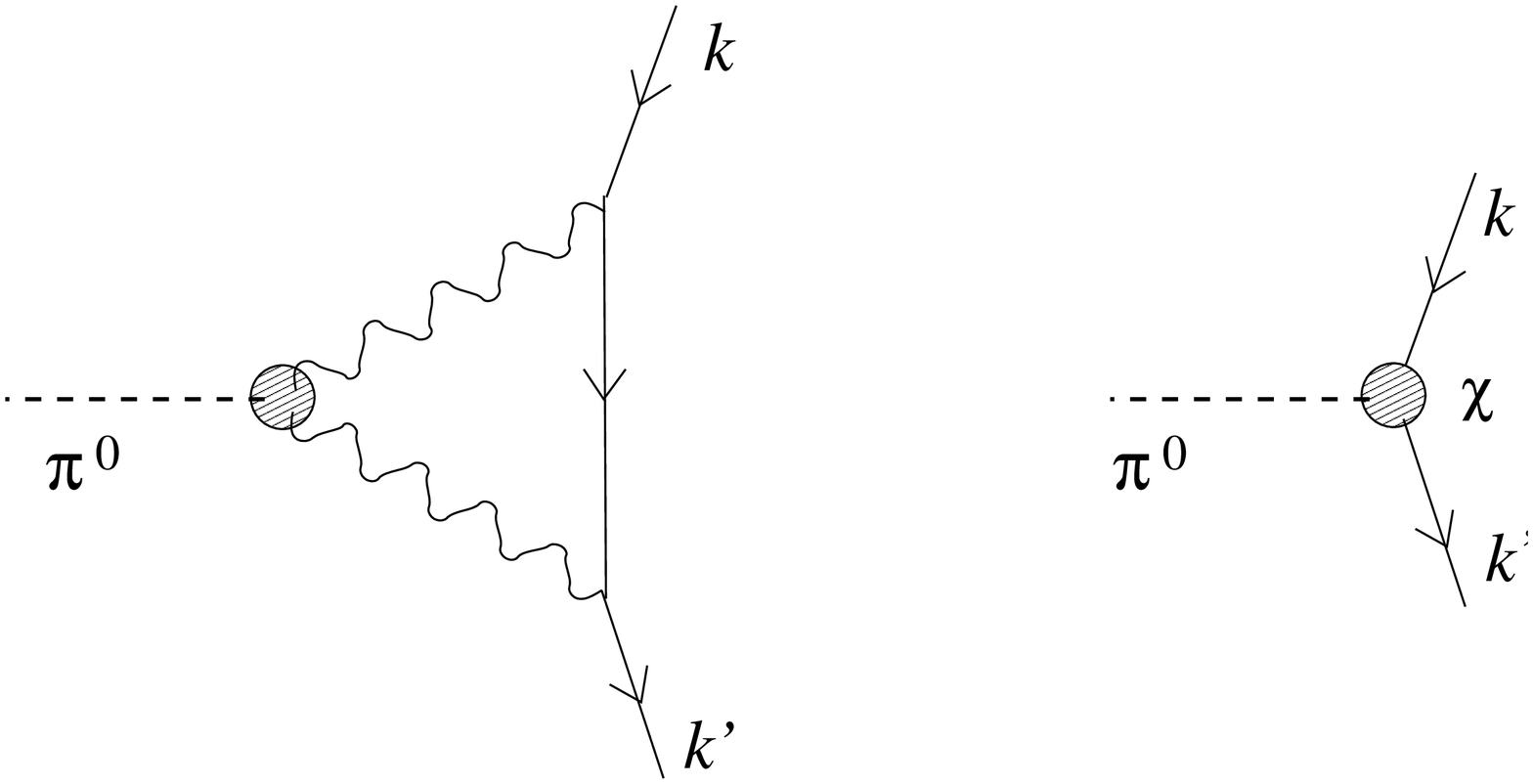,width=10.0cm,height=4.0cm}
\end{center}
\caption{\label{fig:pion2ph}
One loop contribution to the anomalous decay of the neutral pion
into a electron positron pair. Left panel: loop diagram 
from eq.~(\ref{eq:gaugedWZ2}), right panel: contribution from the 
local counterterm eq.~(\ref{eq:counterterm}). There is an analogous 
contribution from the loop diagram with the external electron legs
exchanged, \cf figure~\ref{fig:2ph1}.}
\end{figure}

The second type of diagrams shown is in figure~\ref{fig:2ph1} and has the
photon coupled to the nucleon at a single vertex. Such diagrams are 
curious because they do not appear in simple Dirac theories of the
nucleon which are linear in the covariant derivative. Nevertheless, such 
diagrams are not unknown in hadron physics. In particular, the 
$\pi^0\to \gamma\gamma$ decay induces such a vertex as shown in 
figure~\ref{fig:pion2ph}. In that case the intrinsic structure of 
the $\pi^0\gamma\gamma$ vertex is dictated by the quark triangle 
diagram, \ie the axial anomaly. So we may imagine the nucleon coupling 
to an (off--shell) pion via a Yukawa interaction and the pion 
subsequently decaying into an electron--positron pair as in 
figure~\ref{fig:pion2ph}. As we will observe, this process is negligible 
since (after renormalization) this diagram turns out to vanish in the 
limit $m_{\rm e}\to0$. The reason is that in the interaction Lagrangian the 
two photons couple to the derivative of the pion field and, when computing 
the Feynam diagram, an integration by parts produces a factor $m_{\rm e}$.
Hence the single pion exchange cannot produce a significant contribution to 
electron proton scattering. However, in chiral model multiple pion 
exchanges cease to have derivative couplings to the two photons and 
are hence not necessarily suppressed when $m_{\rm e}\to0$. 

The main purpose of the present investigation is to point out
that this anomaly induced process has a considerable affect for
two photon contributions in the Rosenbluth analysis and that this
process has not been considered previously. Of course, this process 
by itself cannot fully explain the discrepancy to the polarization analysis.

The required anomaly contribution to the local process shown in 
figure~\ref{fig:2ph1} can be perfectly studied within chiral soliton models
for baryons. In these models the chiral field $U$ not only is the 
non--linear representation of the pion field but also describes the
nucleon as a (topological) soliton excitation. In these models we may 
understand the two--photon exchanges shown in figure~\ref{fig:2ph1} as
the coupling of the nucleon's pion cloud to the electron through the
anomaly. We will particularly compute the diagram in figure~\ref{fig:2ph1} 
within the Skyrme model and focus on the anomaly contribution which is
unique because it reflects a QCD property.
Diagrams that have two photons coupled to the nucleon at a single vertex
are unique to chiral models because they are formulated in meson degrees
of freedom. There are also non--anomalous local two--photon couplings. 
These are quite model dependent. Therefore we will here only consider 
the one induced by the non--linear $\sigma$ model as an example and 
observe that its contribution vanishes as $m_{\rm e}\to0$.

The soliton model successfully describes many properties of baryons 
and baryon resonances. Though quantitatively the agreement with empirical data
is on the $\mathcal{O}(1/N_{\rm C})$, \ie 30\%, qualitative aspects are
well explained, as has been recently reviewed in ref.~\cite{Weigel:2008zz}. 
This makes this model a perfect candidate to explore the above posed problem. 
The use of model calculations is the more unavoidable since in due time 
lattice calculations for (off--shell) two--photon form factors will not
be available. Furthermore we recall that in soliton models two--photon 
interactions have previously been successfully considered to compute 
static nucleon polarizabilities~\cite{Polar}. Of course, the current problem 
goes beyond treating these interactions in a static framework.

\section{The model}

We consider the two--flavor Skyrme model for baryons as the simplest
of many models that support the soliton picture. In terms of the non--linear
representation for the isovector pion field $\vec{\pi}$
\be
U={\rm exp}\left[i\frac{\vec{\pi}\cdot\vec{\tau}}{f_\pi}\right]\,,
\label{eq:defU}
\ee
where $\vec{\tau}$ is the vector of Pauli matrices, this
model is defined by the Lagrangian
\be
\mathcal{L}=\frac{f_\pi^2}{4}{\rm tr}
\left(\partial_\mu U\partial^\mu U^\dagger\right)
+\frac{1}{32e_{\rm Sk}^2}{\rm tr}\left(
\left[\partial_\mu U,\partial_\nu U^\dagger\right]
\left[\partial^\mu U,\partial^\nu U^\dagger\right]\right) 
+\frac{f_\pi^2m_\pi^2}{4}{\rm tr}\left(U+U^\dagger-2\right)\,.
\label{eq:Skyrme}
\ee
The leading contribution is the non--linear--$\sigma$--term, which
is supplemented by the Skyrme term which contains four derivatives 
on the chiral field and is required to eventually stabilize the soliton.
Also the chiral symmetry breaking pion mass term is added. The model 
parameters that are determined from meson properties are the pion decay 
constant $f_\pi=93{\rm MeV}$ and the pion mass $m_\pi=138{\rm MeV}$.
On the other hand, the Skyrme term coupling $e_{\rm Sk}\approx 4.0$
may vary within a certain regime imposed by reproducing nucleon static 
properties reasonably well.

The hedgehog configuration that builds the soliton and carries unit baryon 
number reads
\be
U_0(\vec{r})={\rm exp}\left[i\vec{\tau}\cdot\hat{r} F(r)\right]
\label{eq:hedeghog}
\ee
where $r=|\vec{r}|$.
The chiral angle, $F(r)$ is determined from the stationary conditions
that result from \eq{eq:Skyrme} subject to the boundary conditions
$F(0)=\pi$ and $F(\infty)=0$. To generate baryon states with good spin 
and isospin we subsequently introduce 
collective coordinates $A\in SU(2)$ 
\be
U(\vec{r},t)=A(t)U_0(\vec{r})A^\dagger(t)
\label{eq:collcord}
\ee
that parameterize the spin--flavor orientation of the hedgehog and 
quantize them canonically. This turns the time variation of the 
collective coordinates into the spin operator,
\be
\vec{J}=\alpha^2[U_0]\,{\rm tr}\left[(-i)\vec{\tau}A^\dagger\,
\frac{dA}{dt}\right]\,,
\label{eq:spin}
\ee
where the moment of inertia, $\alpha^2[U_0]$ is a functional of the classical 
hedgehog field. Its invariance under combined spin and isospin rotations yields
the isospin operator as 
\be
I_a=-D_{ab}J_b \qquad \mbox{with}\qquad
D_{ab}=\frac{1}{2}\,{\rm tr}\,\left[\tau_aA\tau_bA^\dagger\right]\,.
\label{eq:isospin}
\ee
In this relation $D_{ab}$ refers to the adjoint representation $D_{ab}$ of the 
collective rotations. 

The nucleon wavefunction are Wigner--$D$ functions of the collective coordinates 
in the spin $J=\fract{1}{2}$ and isospin $I=\fract{1}{2}$ representation
\be
\langle A|J=I,t,s\rangle = 
\left[\frac{2J+1}{8\pi^2}\right]^{1/2}
D^{J=I}_{t,s}(A)\,.
\label{eq:bwf}
\ee
Here $|s,t\rangle$ represents a nucleon state with spin and isospin projections
$s=\pm\half$ and $t=\pm\half$, respectively. The identification of total spin 
and isospin originates from the hedgehog structure upon which rotations in 
coordinate and iso-- space are identical.

In practice we obtain operators for 
observables that are expressed in terms of the collective coordinates and their 
time derivatives. We use eqs~(\ref{eq:spin}) and~(\ref{eq:isospin}) to write the 
latter as operators in the space of the collective coordinates and sandwich 
them between the wavefunctions \eq{eq:bwf}. The matrix elements are finally 
obtained as integral over the collective coordinates which are most conveniently
evaluated in terms of their Euler angle representation. Later we will particularly 
require 
\be
\langle s ,t | D_{ab} | s^\prime, t^\prime\rangle =
-\frac{4}{3}\langle s ,t | I_aJ_b | s^\prime, t^\prime\rangle
\qquad \mbox{and}\qquad
\langle s ,t | D_{3a} D_{3b} | s^\prime, t^\prime\rangle =
\frac{1}{3}\langle s ,t | \delta_{ab} | s^\prime, t^\prime\rangle \,.
\label{eq:nmatrix}
\ee
These results and techniques are well established in soliton models and we 
refer to reviews, {\it e.\@g.\@}~\cite{Weigel:2008zz}, for derivation and further 
details. These collective coordinate matrix elements come together with the
matrix elements that emerge from the spatial dependence of the soliton. We will
return to their computation in section IV.

\section{One and two--photon interactions}

We obtain the minimal photon Skyrmion interaction by gauging the Lagrangian. 
Later we will also comment on non--minimal interactions. For the 
local part, \eq{eq:Skyrme} this is straightforwardly
accomplished by replacing the partial derivatives with covariant ones:
\be
\partial_\mu U \longrightarrow D_\mu U= \partial_\mu U -ieA_\mu\left[\hat{Q},U\right]\,.
\label{eq:covder}
\ee
Here $\hat{Q}=\tau_3/2+\ID/6$ and $A_\mu$ are the quark charge matrix and the 
photon field, respectively. Substituting this prescription not only yields
the single photon vertex to the nucleon (represented by the soliton) but also 
higher order interactions, in particular two--photon vertices, the so--called 
seagull terms. For example, the non--linear--$\sigma$--term in \eq{eq:Skyrme} gives
\bea
\mathcal{L}_{{\rm nl}\sigma}^{\rm (gauged)}&=&
\frac{f_\pi^2}{4}{\rm tr}
\left(\partial_\mu U\partial^\mu U^\dagger\right)
-i\frac{f_\pi^2e}{2}A_\mu{\rm tr}\left(\hat{Q}\left[
U^\dagger \partial^\mu U + U\partial^\mu U^\dagger\right]\right)\cr
&&\hspace{2cm}-\frac{f_\pi^2e^2}{4}A_\mu A^\mu\,{\rm tr}\,
\left(\left[\hat{Q},U\right]\left[\hat{Q},U^\dagger\right]\right)\,.
\label{eq:gaugenls}
\eea

The situation is slightly more complicated for the non--local Wess--Zumino 
term~\cite{Wi83} that we do not make explicit because it does not contribute 
to pure hadronic objects in the two--flavor Skyrme model. This term encodes 
the QCD anomaly and yields one and two--photon couplings to the chiral field 
when gauged with respect to the corresponding $U(1)$ group. The techniques to 
compute these couplings are based on a trial and error scenario to obtain a gauge 
invariant quantity. These techniques are widely described in the 
literature~\cite{Wi83,Kaymakcalan:1983qq}
and the result for the gauged Lagrangian is
\bea
\mathcal{L}_{\rm WZ}^{\rm gauged}&=&\frac{e}{16\pi^2}\epsilon^{\mu\nu\rho\sigma}
\Big\{A_\mu {\rm tr} \left(\hat{Q}\left[
U^\dagger\partial_\nu UU^\dagger\partial_\rho UU^\dagger\partial_\sigma U
-U\partial_\nu U^\dagger U\partial_\rho U^\dagger U\partial_\sigma U^\dagger\right]
\right) 
\cr  &&\hspace{0.4cm}
+ieA_\mu \partial_\nu A_\rho {\rm tr} \left(2\hat{Q}^2
\left[U^\dagger\partial_\sigma U-U\partial_\sigma U^\dagger\right]
+\hat{Q}\partial_\sigma U \hat{Q} U^\dagger 
-\hat{Q}U\hat{Q}\partial_\sigma U^\dagger\right) \Big\}\,,
\qquad
\label{eq:gaugedWZ2}
\eea
wherein we substituted the physical value of three color degrees of freedom. 
Most interestingly this contribution to the action generates a vertex for the 
neutral pion to anomalously decay into two photons via the expansion 
$U=1+i\vec{\tau}\cdot\vec{\pi}/f_\pi+\ldots$ of \eq{eq:defU}. Taking into account 
the QED coupling to the electrons, $eA_\mu \overline{\Psi}_e \gamma^\mu \Psi_e$, 
this then describes the decay $\pi^0\to e^{+}e^{-}$ via the Feynman diagram 
displayed in the left panel of figure~\ref{fig:pion2ph}.
This process was exhaustively discussed in ref.~\cite{Savage:1992ac} together with 
its generalization to $\eta\to \mu^{+}\mu^{-}$, etc.\@. As a matter of fact, this 
loop diagram is ultra--violet divergent and induces the counterterm
\be
\mathcal{L}_{\rm c.t.}=
\frac{i\alpha^2}{32\pi^2}\chi(\Lambda) \overline{\Psi}_e \gamma^\mu \gamma_5 \Psi_e
{\rm tr} \left(2\hat{Q}^2
\left[U^\dagger\partial_\sigma U-U\partial_\sigma U^\dagger\right]
+\hat{Q}\partial_\sigma U \hat{Q} U^\dagger 
-\hat{Q}U\hat{Q}\partial_\sigma U^\dagger\right)\,,
\label{eq:counterterm}
\ee
where $\alpha=e^2/4\pi=1/137$ is the QED fine structure constant. Furthermore
$\chi$ is a divergent coefficient\footnote{In ref.~\cite{Savage:1992ac} 
counterterm coefficients were independently introduced for the terms in 
eq.~(\ref{eq:counterterm}). However, gauge invariance enforces them to appear
in exactly the displayed manner.} such that the sum of the two diagrams in
figure~\ref{fig:pion2ph} is finite. Furthermore $\Lambda$ refers to the 
normalization scale that enters to properly reproduce the dimensions of 
loop integrals in dimensional regularization with 
$\int d^4l \to \Lambda^{4-D}\int d^Dl$. We separate the finite, but renormalization
scheme dependent part according to
\be
\chi_{\rm fin}(\Lambda)=
6\left(\frac{4}{4-D}-\gamma+{\rm ln}{4\pi}\right)-\chi(\Lambda)\,,
\label{eq:renorm}
\ee
where $\gamma=0.577\ldots$ is Euler's constant. We consider this renormalization
as effective modeling of an eventual off--shell form factor for the 
$\pi^0\gamma\gamma$ vertex~\cite{Dorokhov:2007bd}. After all, classically 
there is no direct interaction as in \eq{eq:counterterm} between electrons
and hadrons. On the other hand, a vertex form factor that deceases with the 
photon momentum renders the diagram in figure~\ref{fig:pion2ph} finite and thus 
does not induce such a direct (and unphysical) electron pion interaction.

The resulting decay width is most conveniently presented in terms of the 
ratio with respect to the decay into two real photons,
\be
\frac{\Gamma(\pi^0\to e^{+}e^{-})}{\Gamma(\pi^0\to\gamma\gamma)}=
\frac{\alpha^2 m_e^2}{8\pi^2m_\pi^2}\,\frac{\sqrt{\xi^2-1}}{\xi}\,|A(\xi)|^2\,,
\label{eq:width}
\ee
where $\xi=\frac{m_\pi^2}{4m_e^2}$. The complex amplitude $A(\xi)$ has a 
complicated representation in terms of Feynman parameter 
integrals~\cite{Savage:1992ac}. Here it suffices to remark that the renormalization 
scale dependence emerges only through its real part,
\be
\mathsf{Re}A(\xi)=\chi_{\rm fin}(\Lambda)-6\,{\rm ln}\frac{m_\pi^2}{\Lambda^2}
-\left[{\rm ln}\left(\xi^2\right)\right]^2
+2\left[3-2{\rm ln}(2)\right]{\rm ln}\left(\xi^2\right)
+\widetilde{A}(\xi)
\label{eq:renormscale}
\ee
where the reminder, $\widetilde{A}(\xi)$ is independent of $\Lambda$ and finite 
as $m_e\to0$. In this massless limit the ultra--violet finite imaginary part
\be
\mathsf{Im}A(\xi)=\frac{4\pi\xi}{\sqrt{\xi^2-1}}\,
{\rm ln}\left(\xi+\sqrt{\xi^2-1}\right)
\label{eq:ImA}
\ee
also diverges logarithmically.

The empirical datum $(6.3\pm0.5)\times10^{-8}$~\cite{Yao:2006px} for the ratio,
\eq{eq:width} is reproduced in the range $-24<\chi_{\rm fin}(\Lambda)<-10$ when 
the renormalization scale in the second term on the right hand side of 
\eq{eq:renormscale} is set to $\Lambda=1{\rm GeV}$. We will adopt that range 
when investigating two--photon processes in the nucleon sector.

A further remark on the limit $m_e\to0$ is in order. Eventually we want to 
assume this limit when computing the cross section for electron nucleon scattering
since it considerable simplifies the kinematics and, of course, is physically 
meaningful because the energy scales that are involved in this scattering
process are huge compared to the electron mass. 
Obviously the width, \eq{eq:width} vanishes in that limit. This
can be easily understood: $W_\mu$ is a total derivative in the one 
pion approximation. When shuffling this derivative to the electron axial 
current to which the photons in the loop couple, a factor $m_e$ is produced.
So this decay goes together with a helicity flip of the electron. Hence our
renormalization condition prevents us from taking the limit $m_e\to0$ in the 
loop. As can be observed from \eq{eq:renormscale} the loop itself
actually produces a double--logarithmic divergence~\cite{Savage:1992ac}.  

\section{Nucleon form factors}

To compute the transition matrix elements for elastic electron nucleon 
scattering we consider eqs.~(\ref{eq:gaugenls}) and~(\ref{eq:gaugedWZ2})
as perturbation and couple the photons to the electrons as indicated in 
figures~\ref{fig:leading} and~\ref{fig:2ph1}. This procedure is standard 
in QED. However, we also need to compute the nucleon matrix elements for 
the rotating hedgehog configuration, eq.~(\ref{eq:collcord}). These matrix
elements are commonly parameterized in terms of form factors. If we extract
the terms linear in $A_\mu$ and write it as $\mathcal{L}^{(1)}=eJ_\mu A^\mu$
the corresponding matrix elements introduce Dirac and Pauli form factors via
\be
\langle N(\vec{p\,}^\prime)|J_\mu|N(\vec{p})\rangle=
{\overline U}(\vec{p\,}^\prime)\left[\gamma_\mu F_1(Q^2)+
\frac{i\sigma_{\mu\nu}q^\nu}{2M}F_2(Q^2)\right]U(\vec{p})\,,
\label{eq:defff}
\ee
where $p_\mu$ and $p_\mu^\prime$ are the on--shell momenta of the initial
and final nucleons and $q_\mu=p_\mu^\prime-p_\mu$ is the momentum transfer.
The above definition is the standard Lorentz covariant parameterization
of the matrix elements of the conserved electro--magnetic current, in which 
$U(\vec{p})$ is the nucleon Dirac spinor. Note that $M$ is just a parameter 
in this decomposition and refers to the actual nucleon mass rather than the 
model prediction. It is convenient to introduce ``electric" and ``magnetic" 
(so called Sachs) form factors
\be
G_E(Q^2)=F_1(Q^2)-\frac{Q^2}{4M^2}F_2(Q^2)\ , \qquad
G_M(Q^2)=F_1(Q^2)+F_2(Q^2)\,,\qquad
\label{eq:emff}
\ee
that show up in the differential cross section, \eq{eq:Rbluth}.
By pure definition the above form factors concern one--photon 
couplings to the nucleon. Unfortunately, they cannot easily be
accessed from data because nature does not terminate at that
order of perturbation theory.

Similarly the two--photon couplings that we extract 
from eqs.~(\ref{eq:gaugenls}) and~(\ref{eq:gaugedWZ2}), 
\be
\mathcal{L}^{(2)}_{{\rm nl}\sigma}=e^2A_\mu A^\mu S 
\qquad \mbox{and} \qquad
\mathcal{L}^{(2)}_{\rm WZ}=e^2\epsilon^{\mu\nu\rho\sigma}
A_\mu\partial_\nu A_\rho W_\sigma
\label{eq:twogacoup}
\ee
respectively, also define form factors. While there is only a single and
simple Lorentz structure for $S$,
\be
\langle N(\vec{p\,}^\prime)|S|N(\vec{p})\rangle=
{\overline U}(\vec{p\,}^\prime) S_{{\rm nl}\sigma} (Q^2) U(\vec{p})\,,
\label{eq:def2nls}
\ee
the anomaly requires $W_\mu$ to be an axial vector with the 
decomposition
\be
\langle N(\vec{p\,}^\prime)|W_\mu|N(\vec{p})\rangle=
{\overline U}(\vec{p\,}^\prime)\left[\gamma_\mu F_A(Q^2)+
q_\mu F_p(Q^2)+i\sigma_{\mu\nu}q^\nu F_E(Q^2)\right]\gamma_5U(\vec{p})\,.
\label{eq:defWff}
\ee
There is no simplification or relation between these form factors as for the 
ordinary axial current because no conservation law applies to $W_\mu$. Yet we 
will see that for unpolarized scattering, that concerns the Rosenbluth method, 
only the first form factor, $F_A$ contributes.

In the soliton model a major task consists in computing the momentum dependent 
form factors in eqs.~(\ref{eq:defff}), (\ref{eq:def2nls}) and~(\ref{eq:defWff}). 
In principle an additional collective coordinate that parameterizes the
position of the soliton must be introduced in eq.~(\ref{eq:collcord}).
Its conjugate momentum will be the linear nucleon momentum.
This yields quite a simple recipe to handle the linear momentum part of 
the matrix elements~\cite{Braaten:1986md}: just take the
Fourier transformation with respect to (minus) the momentum transfer
of the coordinate dependent factors in the decomposition of the current
operators after substituting the soliton configuration. Essentially 
we will have to fold the radial functions in the currents by spherical
Bessel functions associated with angular momentum of the multiply 
angular structure. In general we may choose any frame to do this
calculation. However, it turns out that the Breit frame with
\be
\vec{p}=-\vec{p\,}^\prime=-\frac{\vec{q}}{2}
\qquad {\rm and}\qquad  q^0=0
\label{eq:breitframe}
\ee
is particularly suited not only because it properly reflects the zero
energy transfer onto an infinitely heavy (large $N_C$) soliton but also
because it directly connects the electric
form factor, $G_E$ and the magnetic form factor, $G_M$ to the time and
spatial components of the electro--magnetic current, respectively.
In this frame the incoming and outgoing nucleons evidently have the 
same energy $E=p^0=p^{\prime 0}=\sqrt{M^2+Q^2/4}$.  For baryons with spin 
$\frac{1}{2}$ we the find the Sachs form factors from the matrix elements 
\bea
\langle N(\vec{p\,}^\prime)|J^0(0)|N(\vec{p})\rangle&=&
2MG_E(Q^2)\langle s_3^\prime |s_3\rangle \cr
\langle N(\vec{p\,}^\prime)|J^i(0)|N(\vec{p})\rangle&=&
-2iG_M(Q^2)\epsilon^{ijk}q^j
\langle s_3^\prime |S_k|s_3\rangle\,,
\label{eq:ffbreitframe}
\eea
where $\vec{S}$ is the nucleon spin operator.

So far we have treated the model in a non--relativistic fashion that 
restricts the energy range to be reliably considered below the nucleon mass.
Using the techniques and results of refs.~\cite{Ji:1991ff,Holzwarth:2005re}
we may extend the nucleon form factor calculation to larger momenta by 
the transformation
\bea
G_E(Q^2)\longrightarrow \gamma^{-2n_E}G_E\left(\frac{Q^2}{\gamma^2}\right)
\qquad{\rm and}\qquad
G_M(Q^2)\longrightarrow \gamma^{-2n_M}G_M\left(\frac{Q^2}{\gamma^2}\right)\,,
\label{eq:ffboost}
\eea
where $\gamma=\sqrt{1+\tau}$ is the Lorentz boost factor. Essentially this 
transforms the from factors from the non--relativistic to a relativistic frame. 
It is worthwhile to note that operator ordering ambiguities in quantizing
the linear momentum are mitigated by choosing the Breit frame as starting point.
This is so because we have $\vec{p\,}^2=\vec{p\,}^{\prime2}$ for this special 
frame. The original study~\cite{Ji:1991ff} is based on the Lorentz boost 
and suggests to put $n_E=0$ and $n_M=1$. For more insight it is instructive 
to reflect on the nature of the transformation, eq.~(\ref{eq:ffboost}).
Most evidently the momentum interval $[0,4M^2]$ of the rest frame is mapped
onto the space--like momenta in the Breit frame.  While small momenta are
almost unaffected, the form factors at infinity in the Breit frame are
obtained from those in the rest frame at $Q^2=4M^2$. Even though the latter
may be small, there is no general reason for them to vanish. In particular,
this implies that the form factors do not match the empirical dipole form
unless $n_E=n_M\ge2$. Thus the values $n_E=n_M\ge2$ are also frequently
adopted because they are strongly motivated by regarding the baryon as a 
cluster of particles whose leading Fock component is a three particle
state~\cite{Mitra:1976cf,Kelly:2002if}. In any case, the large $Q^2$ behavior
is not a profound model result but merely originates from the boost
prescription and thus mainly reflects the kinematical situation. We will 
henceforth assume $n_E=n_M=2$ and similarly $n_{\rm WZ}=2$ for the 
form factors in \eq{eq:defWff}. We note that there is an additional ambiguity 
in the choice of the mass parameter in the Lorentz factor~$\gamma$. We 
take $M$ to be the nucleon mass, yet from the point of view from an $1/N_C$ 
expansion one could equally well argue for the soliton mass which is about
50\% larger. Again, this does not significantly affect qualitative results.

\begin{figure}[t]
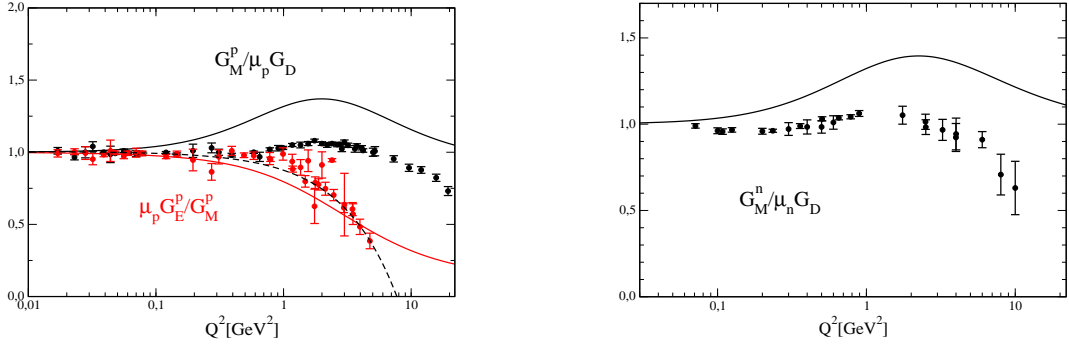

\centerline{
\epsfig{file=gmp.eps,height=4.5cm,width=6.0cm}\hspace{2cm}
\epsfig{file=gmn.eps,height=4.5cm,width=6.0cm}}
\caption{\label{fig:jiff}
Proton (left) and neutron (right) form factors computed in the
Skyrme model as a function of momentum transfer after applying the
boost. These data have been generated with $e_{\rm Sk}=3.8$. The magnetic
from factors are additionally normalized with respect to the predicted 
magnetic moments $\mu_{\rm p}=G^{\rm p}_{\rm M}(0)\approx 2.33$
and $\mu_{\rm p}=G^{\rm n}_{\rm M}(0)\approx-1.99$. The dashed line 
represents the empirical fit, \eq{eq:Rpol}. Data are from 
refs.~\cite{Milbrath:1997de,Jones:1999rz,Pospischil:2001pp,Gayou:2001qt,Gayou:2001qd,Punjabi:2005wq,Crawford:2006rz,Jones:2006kf,Ron:2007vr}.}
\end{figure}

In figure~\ref{fig:jiff} we show the resulting form factors $G_E$ and
$G_M$ and compare them with data. The data for $G_M$ are obtained according
to the Rosenbluth method while the ratio $G_M/G_E$ is taken from polarization
measurements that are assumed to be more robust against the two--photon 
contamination. 
As usual we display these data normalized to the dipole form factor,
\be
G_D(Q^2)=\frac{1}{\left(1+Q^2/0.71{\rm GeV}^2\right)^2}\,.
\label{eq:dipole}
\ee
This figure clearly demonstrates that soliton models are able to reproduce 
the gross features of the empirical form factors. This is particularly the case 
for the linear fall of the ratio $R(Q^2)$, \cf eqs.~(\ref{eq:ratio}) 
and~(\ref{eq:Rpol}). Taking the point of view, that two--photon corrections
are small for the polarization method~\cite{Carlson:2007sp} it is hence 
our task to explain the Rosenbluth cross section within the model.

\begin{figure}[t]
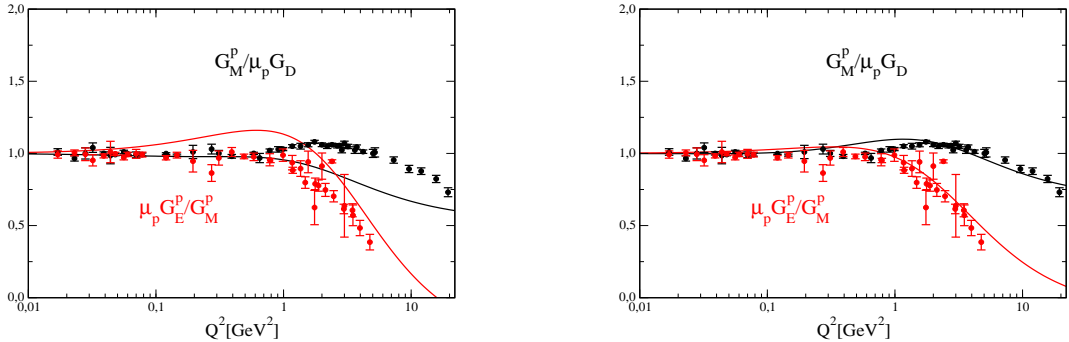

\vskip0.2cm
\centerline{
\epsfig{file=gmpL9full.eps,height=4.5cm,width=6.0cm}\hspace{2cm}
\epsfig{file=gmpL9half.eps,height=4.5cm,width=6.0cm}}
\caption{\label{fig:jiffL9}
The influence of the non--minimal photon coupling, eq.~(\ref{eq:l9term}),
on the proton form factors for two parameters $L_9=0.0069$ (left)
and $L_9=0.0045$ (right). All other model parameters are as in
figure~\ref{fig:jiff}.}
\end{figure}

Nevertheless, deviations of the model predictions from the actual data are 
apparent. Model modifications can improve the agreement with data. For 
example, it is known from chiral perturbation theory studies on the pion 
radius~\cite{Gasser:1984gg} that non--minimal photon couplings as in
\be
{\mathcal L}_9=-iL_9A^{\mu\nu}{\rm tr}\left[Q\left(
\partial_\mu U \partial_\nu U^\dagger +
\partial_\mu U^\dagger \partial_\nu U\right)\right]\,,
\label{eq:l9term}
\ee
are mandatory to correctly reproduce the pion electromagnetic 
properties. In eq.~(\ref{eq:l9term}) $A^{\mu\nu}$ is the photon field 
strength tensor and $L_9\approx0.0069$ is adjusted to the pion 
radius~\cite{Meissner:2007tp}. This term may be understood as resembling 
the contribution from (short distance) vector meson 
fields that have been integrated out when approximating the 
effective chiral theory by the Skyrme model~\cite{Schwesinger:1988af}.
In figure~\ref{fig:jiffL9} we show the proton form factors when the non--minimal 
electro--magnetic coupling, eq.~(\ref{eq:l9term}), is incorporated in the
the electromagnetic current.
As can be seen from the right panel of that figure, a moderate adjustment 
of the additional parameter indeed leads to excellent 
agreement with data. This is particularly the case for the linear 
decay of $R$. Being a total derivative, eq.~(\ref{eq:l9term}) does
not affect the form factors at zero momentum transfer.

The reader may also consult ref.~\cite{Holzwarth:2005re} for a more thorough 
investigation in a vector meson soliton model that strongly supports the above 
statement that soliton models provide a fair account of the nucleon form factors, 
even at large momenta. That investigation also shows that the just mentioned 
deviation from the data can be removed by fine tuning the model. The Lorentz 
boost, \eq{eq:ffboost} is crucial to gain that agreement. As explained in 
ref.~\cite{Holzwarth:2005re} the strong decrease of the ratio $R(Q^2)$
then emerges naturally in chiral soliton models as it basically stems
from the isospin being generated from a rigid rotation in flavor space.
In any event, we are mainly interested in whether or not the two--photon 
exchange coupled via the anomaly significantly contributes to the Rosenbluth 
cross section. To answer this question qualitatively, no further fine 
tuning of the model to reproduce the from factors in detail is required.

\begin{figure}[t]
\begin{center}
\bigskip~
\epsfig{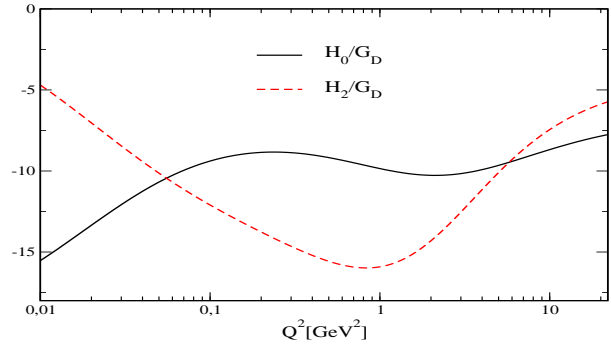}
\end{center}
\caption{\label{fig:h0h2}
Nucleon axial form factors that contribute at the level of the two--photon 
exchange to the elastic electron nucleon scattering. They are computed 
from \eq{eq:H0H2} with the boost, \eq{eq:ffboost} applied with $n_{\rm WZ}=2$ and 
normalized to the dipole form \eq{eq:dipole}. Again we adopted $e_{\rm Sk}=3.8$.}
\end{figure}

As will be shown in the next chapter, only the form factor $F_A$ contributes 
to the interference with the one--photon exchange. For its computation in the
Breit frame we consider the spatial components of $W^\mu$ in eqs~(\ref{eq:gaugedWZ2})
and~(\ref{eq:counterterm}),
\be
\langle N(\vec{p\,}^\prime)|W_i|N(\vec{p})\rangle=\pm
\frac{1}{9\pi M} \chi^{\prime\dagger}\left[H_0(Q^2)\sigma_i
+H_2(Q^2)\left(\sigma_i-3\hat{q}_i \hat{q}\cdot\vec{\sigma}\right)\right]\chi\,.
\label{eq:Wmatrix}
\ee
Here $\chi$ refers to the two--component nucleon spinor. Doting this matrix
element into $\hat{q}$ as well as averaging the $\vec{q}$ directions yields
two relations between $H_i$ and the form factors in \eq{eq:defWff} from which
\be
F_A(Q^2)=\pm \frac{1}{18\pi M E}\left[H_0(Q^2)+H_2(Q^2)\right]
\label{eq:FA}
\ee
is extracted. The two signs refer to proton and nucleon, respectively.
In the next step we compute the left hand side of \eq{eq:Wmatrix} in
the soliton model with the techniques of ref.~\cite{Braaten:1986md}: 
We substitute the rotating hedgehog configuration, \eq{eq:collcord}
into the expression for $W_i$ that we extracted from eqs~(\ref{eq:gaugedWZ2})
and~(\ref{eq:counterterm}) and take matrix elements between nucleon
states. They elements are straightforwardly evaluated with the help of 
\eq{eq:nmatrix} and by noting that 
$A^\dagger\hat{Q}A=\fract{1}{6}\ID+\fract{1}{2}D_{3i}\tau_i$.
Finally we encounter the Fourier transforms of the chiral angle in the form
\bea
H_0(Q^2)&=&M^2\int_0^\infty drr^2\, 
\left[\frac{dF}{dr}+\frac{2}{r}{\rm sin}F{\rm cos}F\right]
j_0(|\vec{q}|r) \cr
H_2(Q^2)&=&M^2\int_0^\infty drr^2\,
\left[\frac{dF}{dr}-\frac{1}{r}{\rm sin}F{\rm cos}F\right]
j_2(|\vec{q}|r)\,,
\label{eq:H0H2}
\eea
where $j_\ell(z)$ denotes the spherical Bessel functions associated with 
orbital angular momentum $\ell$. Once these momentum dependent 
functions are computed they are subject to the boost transformation,
\eq{eq:ffboost} with $n_{\rm WZ}=2$. The resulting form factors $H_0$ and $H_2$
are displayed in figure~\ref{fig:h0h2}.

As long as we omit time derivatives of the collective coordinates, the spin
operator, \eq{eq:spin} does not explicitly enter. In any event, it can only
occur in $W_0$ which is the only component that contains a time derivative.  
Direct evaluation of $\langle W_0\rangle$ gives zero for this nucleon matrix 
element after hermitionizing eventual ordering ambiguities.

\section{Results}

We mainly intend to point out that there is an effect of the anomaly 
that reveals itself as relevant two--photon contributions to electron 
nucleon scattering processes. Hence we do not attempt any fine
tuning of parameters.

To estimate this effect we merely have to substitute the matrix element 
of $W_\mu$ in the form factor decomposition, \eq{eq:defWff} into the Feynman 
diagrams of figure~\ref{fig:2ph1}. We may then formally write the corresponding 
matrix element for the electron nucleon scattering as
\bea
iM_{2\gamma}^{\rm WZ}(q^2)&=&-i\alpha^2{\overline u}(k^\prime)
\left[w_1(q^2)k^{\prime\mu}\gamma_5
+w_2(q^2)q^\mu\gamma_5+w_3(q^2)\gamma^\mu\gamma_5\right]u(k)\cr\cr
&&\hspace{1cm}\times
{\overline U}(\vec{p\,}^\prime)\left[\gamma_\mu F_A(Q^2)+
q_\mu F_p(Q^2)+i\sigma_{\mu\nu}q^\nu F_E(Q^2)\right]\gamma_5\,U(\vec{p})\,.
\label{eq:WZmatrix}
\eea
Again $q_\mu$ is the momentum transferred from the electrons (represented 
by the spinors $u(k)$ and $u(k^\prime)$) to the protons via the two photons.
The above parameterization is general for couplings via axial currents that
enter here because of the $\epsilon$--tensor in the Wess--Zumino 
term~\cite{Rekalo:2003xa}.
Henceforth we will no longer make explicit the dependence on the 
momentum transfer $q^2=-Q^2$, which is negative for this scattering process.

The effects of the photon--electron loop are contained in the form factors 
$w_i(q^2)$. We will discuss the relevant pieces thereof later. We are mainly
interested in the (unpolarized) interference with the one photon exchange
whose transition matrix element is given by the electro--magnetic nucleon 
form factors, \cf \eq{eq:defff},
\bea
iM_{\gamma}&=&i\frac{4\pi\alpha}{q^2}{\overline u}(k^\prime)\gamma^\mu u(k)
{\overline U}(\vec{p\,}^\prime)\left[\gamma_\mu F_1(Q^2)+
\frac{i\sigma_{\mu\nu}q^\nu}{2M}F_2(Q^2)\right]U(\vec{p})\cr
&=&i\frac{4\pi\alpha}{q^2}{\overline u}(k^\prime)\gamma^\mu u(k)
{\overline U}(\vec{p\,}^\prime)\left[\gamma_\mu G_M(Q^2)
-\frac{1}{2M}\left(p_\mu+p^\prime_\mu\right)F_2(Q^2)\right]U(\vec{p})\,,
\label{eq:matrix}
\eea
where in the second equation we employed the Gordon decomposition.
We sum the interference contributions over electron and proton spins. This 
computation is significantly simplified by using of momentum conservation and 
the fact that the spinors $U(\vec{p})$ and $u(\vec{k})$ obey free Dirac equations.
Only a few structures contribute when we average over polarizations. The 
intrinsic parity of $\gamma_5$ requires four additional $\gamma$ matrices 
under the Dirac sum. Two of which stem from 
$U(p){\overline U}(p)=({p \hskip -0.55em /}+m)/2$ and the equivalent construction
from $p^\prime$. A third one comes from the $\gamma_\mu$ in the one photon exchange,
\eq{eq:matrix}. Hence the forth must originate from the from factor decomposition 
in \eq{eq:WZmatrix}. This directly shows that the contribution from the one--photon 
exchange to the interference is linear in the magnetic form factor $G_M$.
Since we also have a Gordon--type decomposition
${\overline U}(p^\prime)\sigma_{\mu\nu}q^\nu\gamma_5U(p)=
i{\overline U}(p^\prime)\left(p_\mu+p^\prime_\mu\right)\gamma_5U(p)$ the Dirac structures
associated with neither $F_p$ nor $F_E$ satisfy this criterion. A similar argument, 
of course, holds for the electron form factors $w_1$ and $w_2$. Finally, the 
intrinsic parity of the $\epsilon$--tensor enforces the 
sum over polarizations to be antisymmetric under $p\leftrightarrow p^\prime$
(or, equivalently $k\leftrightarrow k^\prime$). Up to overall constants, these
considerations determine the final result
\bea
\sum_{\rm spins}M^*_{\gamma}M_{2\gamma}^{\rm WZ}&=&
\frac{128\pi\alpha^3}{q^2}w_3F_AG_M\left[(k\cdot p^\prime)^2-(k\cdot p)^2\right]\cr
&=&128\pi\alpha^3\,w_3F_AG_M M^2
\sqrt{\tau\left(1+\tau\right)\frac{1+\epsilon}{1-\epsilon}}
\label{eq:suminterf}
\eea
that formally shows a deviation from the linear $\epsilon$--dependence found in the 
one--photon exchange approximation, \eq{eq:Rbluth}. We have written this equation such 
as to make explicit the dependence on $\sqrt{(1-\epsilon)/(1+\epsilon)}$ as required 
by general properties and consistency conditions for the two--photon 
interaction~\cite{Rekalo:2003xa}.

If $W_\mu$ were a total derivative, $F_p$ would be the only non--zero 
from factor and thus the interference would vanish in the limit $m_e\to0$.
On the other hand, the $q_\mu$ term of the hadron form factor 
characterizes the $\pi^0\to e^{-}e^{+}$ decay that we discussed earlier.

The information about
the photon loop is contained in the electron form factor that we compute in 
dimensional regularization,
\bea
w_3&=&-2-\int_0^1 dy \int_0^{1-y}dx \,
\frac{\left(1-x\right)q^2}{x^2m_e^2-\left(1-x-y\right)yq^2-i\epsilon} \cr
&& \hspace{1.5cm}
+6\int_0^1 dy \int_0^{1-y}dx \left[\frac{4}{4-D}
-{\rm ln}\left(\frac{x^2m_e^2-\left(1-x-y\right)yq^2}{4\pi\Lambda^2}\right)\right]
-\frac{\chi(\Lambda)}{2}\,. \qquad
\label{eq:elff}
\eea
We have made explicit the contribution from the counterterm, \eq{eq:counterterm} 
that eventually cancels the part that diverges as $D\to4$, according to \eq{eq:renorm} .
It is interesting to consider the leading contribution in the limit of 
vanishing electron mass,
\setlength{\unitlength}{1cm}
\renewcommand{\arraystretch}{0.8}
\be
w_3
\hspace{0.4cm}
\begin{array}{c}
\begin{picture}(0,0)
\put(-0.5,-0.1){\vector(1,0){1.1}}
\end{picture} \cr
{\scriptstyle m_e\to 0}
\end{array}
\hspace{0.5cm}
\widetilde{w}_3=-2{\rm ln}\left(\frac{m_e^2}{Q^2}\right)+7
+\frac{1}{2}\chi_{\rm fin}(\Lambda)-3{\rm ln}\left(\frac{Q^2}{\Lambda^2}\right)
\label{eq:elffm0}
\ee
\renewcommand{\arraystretch}{1.0}\noindent

\noindent
and compare it with \eq{eq:renormscale}. First, we notice that the $\Lambda$--dependence
is the same, so that after fixing the counterterm via the decay $\pi\to e^{-}e^{+}$
the model prediction for the unpolarized cross section does not possess any 
renormalization scale dependence. Second, the $m_e\to0$ singularity is more severe 
for the decay than for the considered cross section. This reflects the fact that
the $\left[{\rm ln}\left(\frac{m_e^2}{m_\pi^2}\right)\right]^2$ divergence is 
buried\footnote{In general the denominator in the first integral in \eq{eq:elff}
always yields a double logarithm as the leading divergence. However, for the
special combination $1-x$ in numerator of the first integral in \eq{eq:elff}
this piece drops out.}
in the electron form factors $w_1$ and $w_2$ that do not contribute to the 
cross section after averaging the polarizations. In turn this implies that for
a prescribed amplitude $A$ in \eq{eq:renormscale}, 
$\chi_{\rm fin}={\mathcal O}\left[{\rm ln}\,m_e^2\right]^2$. Hence, for momenta 
$Q^2\gg m_e^2$ the counterterm would actually dominate this contribution to the 
cross section if we chose such a renormalization condition\footnote{In a fully 
renormalizable theory renormalization conditions are commonly imposed on 
Green's functions with external legs amputated. This would translate to 
constrain the off--shell amplitude $A$ rather than the physical on--shell
decay width $\Gamma$ via a renormalization condition.}. The discussion of
this singularity in ref.~\cite{Dorokhov:2007bd} suggests that it arises 
independently of the high momentum treatment. In ref.~\cite{Afanasev:2004hp} 
the authors carefully analyze the (double)--logarithmic singularities in 
the box diagrams of figure~\ref{fig:box}. In that case, single 
logarithmic singularities emerge always while double logarithmic singularities
only occur when the momentum transfer roughly equals the mass of the 
exchanged hadronic resoncance. The triangle diagrams 
(figure~\ref{fig:2ph1}) do not contain hadronic resonances. Thus the 
non--existence of double logarithmic singularities in their contribution to
electron nucleon scattering is expected; even though they may generally 
emerge as the rare $\pi^0$ decay exemplifies.

To facilitate the discussion of our numerical results for the two--photon exchange 
contribution from the Wess--Zumino term we identify the tree level cross--section, 
\eq{eq:Rbluth} and introduce the reduced cross--section,
\be
\left(\frac{d\sigma}{d\Omega}\right)_R=\frac{\epsilon}{\tau}
\left(1+\tau\right)\frac{d\sigma}{d\Omega}
\Big/
\left(\frac{d\sigma}{d\Omega}\right)_{\rm Mott}\,,
\label{eq:sigmaRdef}
\ee
so that
\bea
\left(\frac{d\sigma}{d\Omega}\right)_R&=&G_M^2+\frac{\epsilon}{\tau}G_E^2
+\frac{2\alpha}{\pi}\widetilde{w}_3 M^2 F_A G_M 
\sqrt{\tau(1+\tau)\left(1-\epsilon^2\right)}\cr
&=&\frac{G_M^2}{\tau} \left[\tau+\epsilon \frac{G_E^2}{G_M^2}
+\frac{2\alpha}{\pi}\widetilde{w}_3 M^2 \frac{F_A}{G_M} 
\sqrt{\tau^3\left(1+\tau\right)\left({1-\epsilon^2}\right)}\right] \,.
\label{eq:sigmaR}
\eea
For small $\epsilon$ the predicted correction to the Rosenbluth form is 
${\mathcal O}\left(\epsilon^2\right)$ and thus small. Note, that in the form 
of the second equation the ambiguities in choosing the powers $n_X$, 
\cf \eq{eq:ffboost} cancel within the square bracket.

\begin{figure}[t]
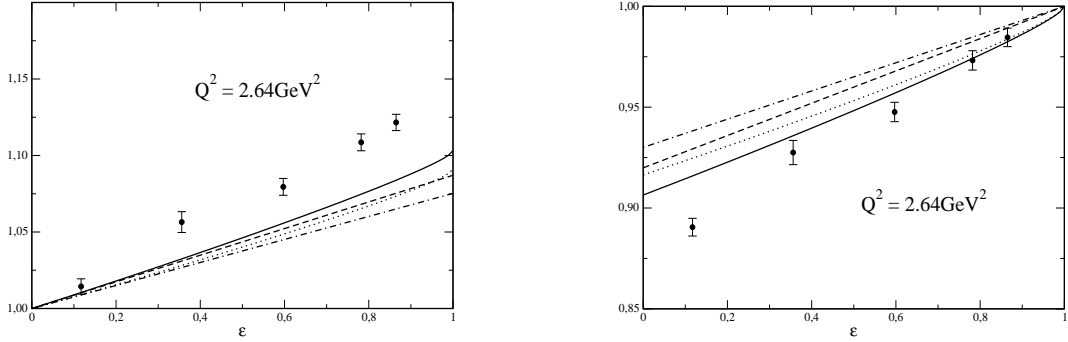

\centerline{
\epsfig{file=f1f4at0.eps,height=4.5cm,width=6.0cm}\hspace{2cm}
\epsfig{file=f1f4at1.eps,height=4.5cm,width=6.0cm}}
\caption{\label{fig:f1f4}
Results for the normalized reduced cross section, \eq{eq:rmodel},
compared to experimental data~\cite{Qattan}. The full line is our model result,
the dashed line is the model result at the one--photon exchange
level, \ie $F_A\equiv0$. 
The dotted and dashed--dotted lines are similarly obtained
with the polarization result for $G_E/G_M$, \eq{eq:Rpol} substituted
for the second term of the numerator in \eq{eq:rmodel}. The left and 
right panels distinguish the normalization
at $\epsilon=0$ and $\epsilon=1$, respectively.}
\end{figure}

\begin{figure}[t]
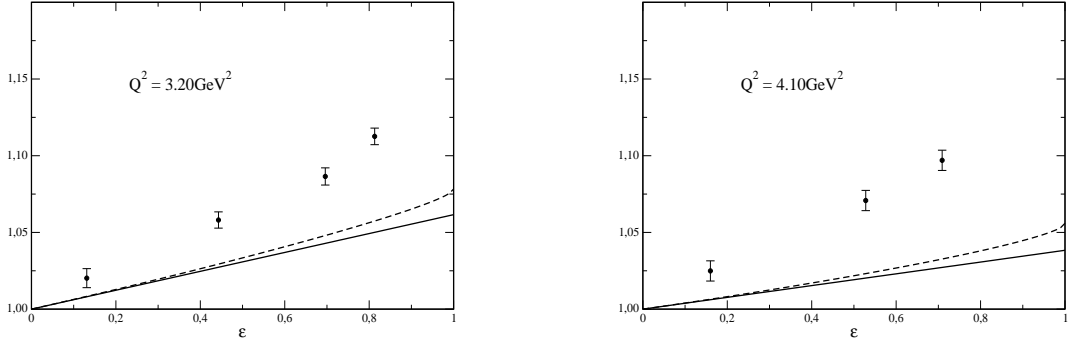

\bigskip
\centerline{
\epsfig{file=f1f2w320.eps,height=4.5cm,width=6.0cm}\hspace{2cm}
\epsfig{file=f1f2w410.eps,height=4.5cm,width=6.0cm}}
\caption{\label{fig:f1f2}
Same as the left panel of figure~\ref{fig:f1f4} for two different 
values of $Q^2$.}
\end{figure}

\begin{figure}[t]
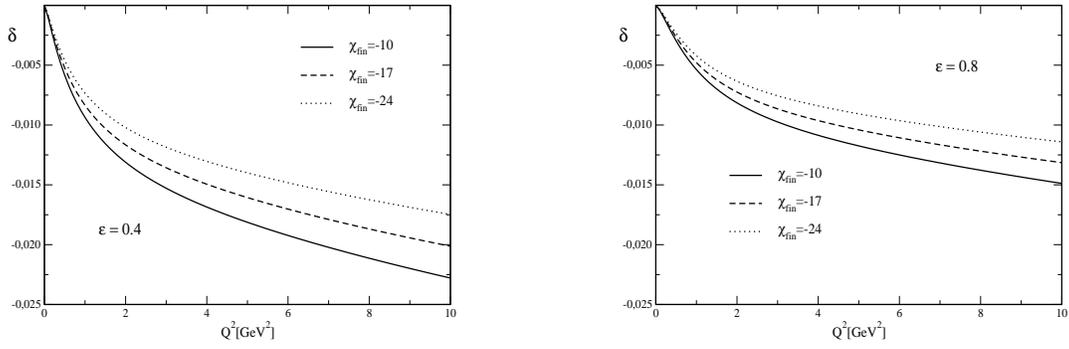

\vskip0.4cm
\centerline{
\epsfig{file=del04.eps,height=4.5cm,width=6.0cm}\hspace{2cm}
\epsfig{file=del08.eps,height=4.5cm,width=6.0cm}}
\caption{\label{fig:delta}
The two--photon exchange contribution, $\delta$ in \eq{eq:defdelta}
for two values of the photon polarization parameter $\epsilon$ as a 
function of the momentum transfer $Q^2$. Also given are the results
for different finite parts of the counterterm coefficient, \eq{eq:renorm}.}
\end{figure}

In figures~\ref{fig:f1f4} and~\ref{fig:f1f2} 
we present our result for $\left(\frac{d\sigma}{d\Omega}\right)_R$.
To mitigate the model deficiencies associated with $G_M$ we compare the model 
prediction 
\be
r_{\rm mod}(\epsilon)=
\frac{1+\frac{\epsilon}{\tau} \frac{G_E^2}{G_M^2}
+\frac{2\alpha}{\pi}\widetilde{w}_3 M^2 \frac{F_A}{G_M}
\sqrt{\tau\left(1+\tau\right)\left({1-\epsilon^2}\right)}}
{1+\frac{2\alpha}{\pi}\widetilde{w}_3 M^2 \frac{F_A}{G_M}
\sqrt{\tau\left(1+\tau\right)}}
\label{eq:rmodel}
\ee
to the ratio of empirical data
\be
r_{\rm exp}(\epsilon)=\frac{(d\sigma(\tau,\epsilon)/d\Omega)_R}
{(d\sigma(\tau,0)/d\Omega)_R}\,.
\label{eq:rexp}
\ee
By construction, the experimental value $(d\sigma(\tau,0)/d\Omega)_R$ is the 
magnetic form factor as obtained from the Rosenbluth method.
We recognize from these figures that the additions from the Wess--Zumino 
term to the unpolarized cross section work into the direction required by the 
data. However, they are about a factor five too small for low $Q^2$. 
For larger $Q^2$ it may fall short by an order of magnitude.
When we normalize with respect to $\epsilon=0$ these additions 
are strongest around the end--point $\epsilon\to1$. One the other hand,
this normalization point is not very special and we may adopt equally 
well $\epsilon=1$, as displayed in the right panel of figure~\ref{fig:f1f4}. 
In that case the agreement with data occurs to be significantly better, 
yet it is merely a matter of presentation. It also suggests that this 
two--photon effect would be most strongly pronounced around $\epsilon\approx0$.
This is, of course, not the case as can easily be recognized by
inspecting \eq{eq:suminterf}. However, a common procedure in the 
literature, that discusses the two--photon contamination in terms of 
the quantity $\delta$ defined via
\be
\left(\frac{d\sigma}{d\Omega}\right)_R
=\left[G_M^2+\frac{\epsilon}{\tau}G_E^2\right](1+\delta)
\label{eq:defdelta}
\ee
and that we show in figure~\ref{fig:delta}, suggests otherwise because of 
the $\epsilon$ dependence of the pre--factor. Nevertheless such a 
presentation is interesting as it disentangles the two--photon exchange contribution,
\ie the last term in \eq{eq:sigmaR} normalized to the one--photon contribution
to the cross section. Experimentally this corresponds to the ratio of the 
difference and the sum of the cross sections for unpolarized electron--proton 
and positron--proton scattering. There are no new data on this separation.
However, the existing data~\cite{Browman:1965aa,Mar:1968qd} 
indicate that $\delta$ should be negative~\cite{Blunden:2005ew}.

\begin{figure}[t]
~\vskip0.2cm
\centerline{
\epsfig{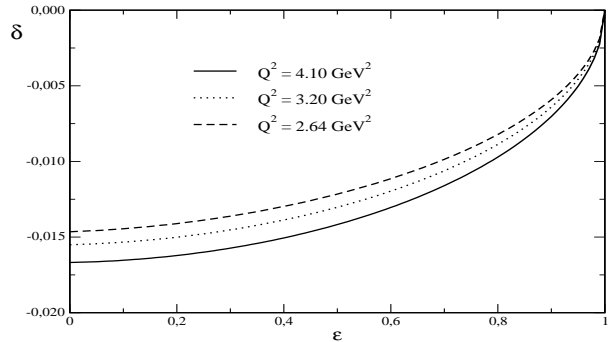}}
\caption{\label{fig:feps}
The model prediction for the two--photon exchange contribution, $\delta$ 
as a function of the photon polarization parameter $\epsilon$.}
\end{figure}

We observe that the anomaly contribution to $\delta$ has a large 
slope at small momentum transfer, $Q^2$ while it levels off with
increasing $Q^2$. This dependence is uniform as we vary~$\epsilon$.
As a function of $\epsilon$ with fixed $Q^2$ we find the largest
slope of $\delta$ around $\epsilon\sim1$, as shown in figure~\ref{fig:feps}.
These features are quite different from the contributions of the 
box diagrams (figure~\ref{fig:box}) that were estimated in 
ref.~\cite{Blunden:2005ew} outside of any soliton model. We stress 
that this is not a contradiction, rather the opposite is the case 
because these results must be added in a full computation of the 
two--photon effects onto the Rosenbluth method.

After \eq{eq:ImA} we argued that we set the renormalization scale
$\Lambda=1{\rm GeV}$ and considered three cases 
$\chi_{\rm fin}(\Lambda)=-24,-17,-10$ that are suggested by the data for the 
decay $\pi^0\to e^{-}e^{+}$. We display the corresponding variations
for the cross section in figure~\ref{fig:delta}. This uncertainty translates 
into an 10--20\% effect for the two--photon contribution to the cross section.
These minor variations with the uncertainty in the fixing the model parameter 
from the underlying process $\pi^0\to e^{+}e^{-}$ is reassuring as it shows that 
the ${\rm ln}\,m_e^2$ effects are not as severe as suspected. All results shown 
in figures~\ref{fig:f1f4} and~\ref{fig:f1f2} refer to the central value 
$\chi_{\rm fin}(\Lambda)=-17$.

We have also investigated the non--anomalous two--photon vertex that 
originates from the non--linear $\sigma$--model, \eq{eq:twogacoup}.
Since this interaction does not have any derivative operator, the 
corresponding triangle diagram is ultra--violet finite and thus 
no counterterm is required. Yet we find that the corresponding matrix 
element, $M_{2\gamma}^{{\rm nl}\sigma}$ vanishes as $m_e\to0$. Hence
this interaction gives negligible contribution within the Rosenbluth 
method, if at all.

\section{Conclusion}

We have performed a model calculation to shed some light on the 
discrepancies that arise from different methods to extract the 
nucleon electro--magnetic form factors from data. These discrepancies
are assumed to be resolved by the inclusion of two--photon processes
in the computation of electron--proton reactions.  Here we have focused 
on the contribution of such a process with the least model dependence
and fewest assumptions about off--shell form factors. This appears to be the
anomaly induced two--photon vertex because it actually is a QCD property. 
It naturally emerges from the nucleon pion cloud coupling to the QCD 
anomaly. This particular two--photon exchange contribution to the elastic
electron nucleon scattering has previously been overlooked presumably 
because it vanishes in the one--pion exchange approximation. However there
is no reason for it to be particularly small beyond that approximation.
For these reasons we focus on this particular process, which of course
does not fully explain the observed discrepancies by itself. Nevertheless,
it is interesting to investigate this effect by its own as it has not been 
considered previously in the context of electron nucleon scattering. Of
course, there is no reason to assume that this piece by itself fully 
resolves the discrepancy between the Rosenbluth and polarization analyses.
At face value the corresponding Feynman diagram is ultra--violet
divergent and requires renormalization. We impose a renormalization 
condition that reproduces the empirical decay width for the process
$\pi^0\to\gamma\gamma$. This is an {\it ad hoc} attempt to deal with 
the (unknown) off--shell behavior of the anomalous $\pi^0\to\gamma\gamma$
interaction that has been successfully utilized for the description of the 
pion decay. The Skyrme soliton model is a perfect and the simplest tool to 
study this anomalous contribution to electron proton scattering because it 
provides both, the pion cloud picture of the nucleon and a unique description 
of the pion anomaly coupling via the Wess--Zumino term. We do not 
exclude that more sophisticated models might provide more reliable 
estimates of this effect.

The so--computed anomaly contribution to the unpolarized cross section
has a minor effect on the cross section, of the order of a few per cent. This is 
to be anticipated for an order $\alpha=1/137$ contamination. Even though this
contribution corrects the leading order result into the proper direction 
these corrections are not sufficient to fully explain the observed discrepancy. 
In this context we stress that this anomaly contribution must
be considered in addition to contributions from the box diagrams in 
figure~\ref{fig:box}. Unfortunately, their computation is quite 
model dependent thereby leading to quite some uncertainties, in particular
at large momentum transfers. Eventually they can be reduced somewhat by 
phenomenological input for the generalized parton distributions from the 
amplitude of deeply virtual Compton scattering~\cite{Chen:2004tw}. The studies 
of ref.~\cite{Blunden:2005ew} indicate that these box diagrams are the most 
significant for $\delta$ (the two--photon piece in the unpolarized cross section) 
at small $\epsilon$. The anomaly contribution that we have computed enhances 
$\delta$ at moderate $\epsilon$ so that we expect a negative, say about 5\%, 
effect at small and moderate $\epsilon$ while at the boundary $\epsilon\to1$ 
the contributions of both, the anomaly and the box diagrams are compatible
with zero.

In general the model contains additional triangle diagram (non--anomalous) type
two--photon processes like those shown in figure~\ref{fig:2ph1} where the two photons 
couple simultaneously to the pion cloud of the nucleon. In the present model it
is natural to assume that the dominant such process stems from the non--linear 
$\sigma$ model. We have seen that it vanishes as the momentum transfer is large 
compared to the electron mass.

In the next step we will have to investigate the anomaly contribution in the 
framework of the polarization method. In particular the effects of the form 
factors $F_E$ and $F_p$, that do not show up in the Rosenbluth method, will
be of future interest.

\section*{Acknowledgments}
The authors gratefully acknowledge many fruitful discussions with G.~Holzwarth 
and H.~Walliser. This work is supported in parts by DFG under 
contract--no. We~1254/13--1.

\end{document}